\documentclass[epj]{svjour}
\usepackage{graphics}
\usepackage{url}
\usepackage{color}
\usepackage{mathrsfs}
\usepackage{amssymb}
\usepackage{bm}
\usepackage[numbers]{natbib}
\usepackage{rotating}
\usepackage[normalem]{ulem}
\usepackage{epstopdf}
\usepackage{lipsum}
\usepackage{mathtools}
\usepackage{cuted}
\usepackage{verbatim}
\usepackage[colorlinks,citecolor=blue,linkcolor=blue,anchorcolor=blue,filecolor=blue,
urlcolor=blue]{hyperref}
\usepackage{enumitem}
\usepackage{multirow}

\begin{document}

\title{Neutron star crustal properties from relativistic mean-field models and bulk parameters effects}

\author{M. Dutra, C. H. Lenzi, W. de Paula, and O. Louren\c{c}o}  

\institute{
Departamento de F\'isica, Instituto Tecnol\'ogico de Aeron\'autica, DCTA, 12228-900, 
S\~ao Jos\'e dos Campos, SP, Brazil}

\date{\today}

\abstract{
We calculate crustal properties of neutron stars, namely, mass ($M_{\rm crust}$), radius ($R_{\rm crust}$) and fraction of moment of inertia ($\Delta I/I$) from parametrizations of hadronic relativistic mean-field (RMF) model consistent with symmetric and asymmetric nuclear matter constraints, as well as some stellar boundaries. We verify which one are also in agreement with restrictions of $\Delta I/I \geqslant 1.4\%$ and $\Delta I/I \geqslant 7\%$ related to the glitching mechanism observed in pulsars, such as the Vela one. The latter constraint explains the glitches phenomenon when entrainment effects are taken into account. Our findings indicate that these parametrizations pass in the glitching limit for a neutron star mass range of $M\leqslant 1.82M_\odot$ ($\Delta I/I \geqslant 1.4\%$), and $M\leqslant 1.16M_\odot$ ($\Delta I/I \geqslant 7\%$). We also investigate the influence of nuclear matter bulk parameters on crustal properties and find that symmetry energy is the quantity that produces the higher variations on $M_{\rm crust}$, $R_{\rm crust}$, and~$\Delta I/I$. Based on the results, we construct a particular RMF parametrization able to satisfy $\Delta I/I \geqslant 7\%$ even at $M=1.4M_\odot$, the mass value used to fit data from the softer component of the Vela pulsar X-ray spectrum. The model also presents compatibility with observational data from PSR J1614-2230, PSR J0348+0432, and MSP J0740+6620 pulsars, as well as, with data from the Neutron Star Interior Composition Explorer (NICER) mission.
}


\authorrunning{M. Dutra {\it et al.}}
\titlerunning{Crustal properties from relativistic mean-field models and bulk parameters effects}

\maketitle

\section{Introduction}
\label{intro}

There are at least two approaches widely used to describe many-body systems in which their constituents are submitted to the nuclear interaction. One of them is based on the construction of potentials where free parameters are adjusted in order to reproduce experimental data from known few-nucleon systems, such as the deuteron: a proton-neutron pair of spin~1 and isospin~0 presenting binding energy around 2~MeV~\cite{lig1,lig2}, electric quadrupole momentum of 2.82~mb~\cite{quadrupolo}, and magnetic momentum of 0.86~mn (\cite{mag}). Since the nucleon-nucleon potential is determined, many-nucleon system is obtained, for instance, from Brueckner-Hartree-Fock~\cite{prc3,bethe} method. In contrast, another approach makes use of phenomenological hadronic models based on the mean-field approximation with their free coupling constants directly fitted to reproduce quantities from many-body nuclear systems, such as those from finite nuclei such as binding energy and charge radius of different nuclei, and those from infinite symmetric and asymmetric nuclear matter, such as bulk parameters evaluated at the saturation density $\rho_0$, namely, incompressibility, skewness parameter, symmetry energy, symmetry energy slope, and other. Among nonrelativistic models used in this latter approach, one can cite Skyrme~\cite{sky1,sky2,sky3,sky4,stoneskyrme}, Gogny~\cite{gogny1,gogny2,gogny3,gogny4,gogny6,gogny7,gogny8,gognyic}, Momentum-dependent interaction~\cite{mdi1,mdi2,mdi3,mdi4}, Michigan three-range Yukawa~\cite{m3y1,m3y2,m3y6}, and Simple Effective Interaction~\cite{behera98,behera15,behera13,behera16} models. For a relativistic treatment, on the other hand, the most used models are different version of the so called relativistic mean-field (RMF) models~\cite{mdi3,walecka,PRC90-055203,prcrmf}, represented by a Lagrangian density from which all thermodynamic quantities are derived.

An example of many-nucleon systems described by the aforementioned hadronic models, in addition to finite nuclei and infinite nuclear matter, is the composing matter of some astrophysical objects as neutron stars, for instance. These objects are formed by protons, neutrons, leptons, and other exotic particles, interacting in such a way to ensure the $\beta$-equilibrium condition. According to several theoretical studies, the interior of a neutron star is composed by a solid crust, at low-density, surrounding a liquid homogeneous core at several times $\rho_0$. The crust, estimated to contain around~$1\%$ percent of the total mass of the star, has a complex structure and is extremely important for the understanding of some astrophysical observations~\cite{BPS-1971a,BPS-1971b,Chamel-2008,Pethick-1995a,Pethick-1995b,Lattimer-2000,Lattimer-2001,Lattimer-2007,Steiner-2005} such as X-rays bursts~\cite{Duncan-1998}, and the abrupt spin-up in the rotational frequency of pulsars~\cite{Link-1999,Ho-2015}. 

Pulsars are very stable rotating neutron stars with a period ranging between $1.4$~ms, as shown the recent observation of the pulsar J1748-2446ad~\cite{Hessels-2006}, and some seconds. The period of these objects has a precision close to atomic clocks. However, the timing behavior of these astrophysical objects can be interrupted by sudden jumps, so-called glitches, in their rotational frequency. This phenomenon has been observed in radio pulsars~\cite{Lyne-book,Kaspi-2000} and is explained from the analysis of the superfluid vortices present in the inner crust of the star. The current understanding for the origin of such glitches is the one proposed in Refs.~\cite{nature1,nature2}. The neutron superfluid vortices pinned to the lattice structure of nuclei contained in the inner crust rotate faster than the compact object, since the star itself loses rotational energy due to the electromagnetic radiation emission (radio waves in the case of radio pulsars). As the difference between the vortices and star angular frequencies becomes higher enough, an abrupt unpinning of some vortices takes place and their outward move is verified. By conservation of angular momentum, the star increases its rotation and the glitch is then established. Some theoretical studies show that pulsar glitches are related to the crustal fraction of the moment of inertia, $\Delta I/I$ ($I$ is the total moment of inertia of the star)~\cite{Atta-2017,Madhuri-2017,Margaritis}. In Ref.~\cite{Link-1999}, authors used observational data from the most active glitching pulsar, the Vela one, and other six pulsars to obtain the glitch constraint of $\Delta I/I \gtrsim 1.4\%$. On the other hand, in Refs.~\cite{PRL109-241103,PRL110-011101}, a higher estimation for this constraint was proposed, namely, $\Delta I/I \gtrsim 7\%$. The difference between the values comes from the consideration, in the latter analysis, of non-dissipative neutron-proton scattering effects (crustal  entrainment).

In this work we use parametrizations of the RMF model previously shown to be consistent with symmetric and asymmetric nuclear matter constraints~\cite{PRC90-055203}, as well as some stellar boundaries~\cite{prcrmf,PRC93-025806}, in order to calculate crustal properties of a neutron star, and verify which one of them are also able to describe the limits on $\Delta I/I$ related to glitch mechanism observed in pulsars such as the Vela one~\cite{pulsars, pulsarsb}. We also provides limits for the Vela radius predicted by these parametrizations. Then, we analyze how the bulk parameters of nuclear matter affect the crustal properties of neutron stars, such as the mass enclosed by the crust, the crust thickness and $\Delta I/I$. Specifically, in Sec.~\ref{rmf} we present the mainly equations concerning the RMF model used in our analysis. The neutron star crustal properties calculated from the parametrizations of this model are shown in Sec.~\ref{crustal} along with the study on the influence of the bulk parameters. Finally, in Sec.~\ref{conclusions}, a short summary and our concluding remarks are presented.

\section{Relativistic hadronic model}
\label{rmf}

The original relativistic mean-field model was developed in 1974 by \cite{walecka2}. In this model, based on quantum filed theory, nucleons are described by the Dirac spinor $\psi$, and the exchanged mesons by the scalar and vector fields $\sigma$ and $\omega_\mu$, responsible by attractive and repulsive nuclear interaction, respectively, in symmetric matter. In order to take into account also the isospin asymmetry (different numbers of protons and neutrons), the inclusion of the $\rho$ meson, represented by the isovector field $\vec{\rho}_\mu$, is also needed. Here we study parametrizations of a generalized version of the Walecka model. The fundamental quantity that describes this improved model is the Lagrangian density given by \cite{mdi3} and \cite{PRC90-055203}
\begin{align}
\mathcal{L} &= \overline{\psi}(i\gamma^\mu\partial_\mu - M_{\mbox{\tiny nuc}})\psi 
+ g_\sigma\sigma\overline{\psi}\psi 
- g_\omega\overline{\psi}\gamma^\mu\omega_\mu\psi
\nonumber \\ 
&- \frac{g_\rho}{2}\overline{\psi}\gamma^\mu\vec{\rho}_\mu\vec{\tau}\psi
+\frac{1}{2}(\partial^\mu \sigma \partial_\mu \sigma - m^2_\sigma\sigma^2)
- \frac{A}{3}\sigma^3 - \frac{B}{4}\sigma^4 
\nonumber\\
&-\frac{1}{4}F^{\mu\nu}F_{\mu\nu} 
+ \frac{1}{2}m^2_\omega\omega_\mu\omega^\mu 
+ \frac{C}{4}(g_\omega^2\omega_\mu\omega^\mu)^2 -\frac{1}{4}\vec{B}^{\mu\nu}\vec{B}_{\mu\nu} 
\nonumber \\
&+ g_\sigma g_\rho^2\sigma\vec{\rho}_\mu\vec{\rho}^\mu
\left(\alpha_2+\frac{1}{2}\alpha'_2g_\sigma\sigma\right) 
+ \frac{1}{2}\alpha'_3g_\omega^2 g_\rho^2\omega_\mu\omega^\mu
\vec{\rho}_\mu\vec{\rho}^\mu
\nonumber\\
&+ \frac{1}{2}m^2_\rho\vec{\rho}_\mu\vec{\rho}^\mu
+ g_\sigma g_\omega^2\sigma\omega_\mu\omega^\mu
\left(\alpha_1+\frac{1}{2}\alpha'_1g_\sigma\sigma\right),
\label{lomegarho}
\end{align}
in which $F_{\mu\nu}=\partial_\nu\omega_\mu-\partial_\mu\omega_\nu$ and $\vec{B}_{\mu\nu}=\partial_\nu\vec{\rho}_\mu-\partial_\mu\vec{\rho}_\nu- g_\rho (\vec{\rho}_\mu \times \vec{\rho}_\nu)$. The nucleon rest mass is $M_{\rm nuc}$, and the mesons masses are $m_\sigma$, $m_\omega$, and $m_\rho$. The free coupling constants of the model are given by $g_\sigma$, $g_\omega$, $g_\rho$, $A$, $B$, $C$, $\alpha_1$, $\alpha_1'$, $\alpha_2$, $\alpha_2'$ and~$\alpha_3'$. The use of the field equations for this model, obtained through the Euler-Lagrange equations, along with the mean-field approximation for the fields, allows the determination of the energy density and the pressure of the system. These thermodynamic quantities are written in terms of the energy-momentum tensor as $\epsilon_{\mbox{\tiny had}}=\left<T_{00}\right>$ and $p_{\mbox{\tiny had}}=\left<T_{ii}\right>/3$. 

The respective expressions are given by
\begin{align}
&\epsilon_{\mbox{\tiny had}} = \frac{1}{2}m^2_\sigma\sigma^2 
+ \frac{A}{3}\sigma^3 + \frac{B}{4}\sigma^4 - \frac{1}{2}m^2_\omega\omega_0^2 
- \frac{C}{4}(g_\omega^2\omega_0^2)^2 
\nonumber\\
&+ g_\omega\omega_0\rho
- \frac{1}{2}m^2_\rho\bar{\rho}_{0(3)}^2
+\frac{g_\rho}{2}\bar{\rho}_{0(3)}\rho_3
- \frac{1}{2}\alpha'_3g_\omega^2 g_\rho^2\omega_0^2\bar{\rho}_{0(3)}^2
\nonumber\\
&- g_\sigma g_\omega^2\sigma\omega_0^2
\left(\alpha_1+\frac{\alpha'_1g_\sigma\sigma}{2}\right) 
- g_\sigma g_\rho^2\sigma\bar{\rho}_{0(3)}^2 
\left(\alpha_2+\frac{\alpha'_2 g_\sigma\sigma}{2}\right) 
\nonumber \\
&+ \epsilon_{\mbox{\tiny kin}}^p + \epsilon_{\mbox{\tiny kin}}^n,
\label{denerg}
\end{align}
and
\begin{align}
&p_{\mbox{\tiny had}} = - \frac{1}{2}m^2_\sigma\sigma^2 
- \frac{A}{3}\sigma^3 - \frac{B}{4}\sigma^4 + \frac{1}{2}m^2_\omega\omega_0^2 
+ \frac{C}{4}(g_\omega^2\omega_0^2)^2 
\nonumber\\
&+ g_\sigma g_\omega^2\sigma\omega_0^2
\left(\alpha_1 + \frac{\alpha'_1g_\sigma\sigma}{2}\right) 
+ g_\sigma g_\rho^2\sigma\bar{\rho}_{0(3)}^2 
\left(\alpha_2+\frac{\alpha'_2g_\sigma\sigma}{2}\right) 
\nonumber \\
&+ \frac{1}{2}m^2_\rho\bar{\rho}_{0(3)}^2
+ \frac{1}{2}{\alpha_3}'g_\omega^2 g_\rho^2\omega_0^2\bar{\rho}_{0(3)}^2
+ p_{\mbox{\tiny kin}}^p + p_{\mbox{\tiny kin}}^n
\label{pressure}
\end{align}
with
\begin{eqnarray}
\epsilon_{\mbox{\tiny kin}}^{p,n} = \frac{\gamma}{2\pi^2}\int_0^{{k_F}_{p,n}}k^2
(k^2+M^{*2})^{1/2}dk
\end{eqnarray}
and
\begin{eqnarray}
p_{\mbox{\tiny kin}}^{p,n} = 
\frac{\gamma}{6\pi^2}\int_0^{{k_F}_{p,n}}\frac{k^4dk}{(k^2+M^{*2})^{1/2}},
\end{eqnarray}
where ${k_F}_{p,n}$ is the proton/neutron Fermi momentum. $\sigma$, $\omega_0$ (zero component) and $\bar{\rho}_{0_{(3)}}$ (isospin space third component) are the expectation values of the mesons fields in the expressions above. The effective nucleon mass is $M^*= M_{\mbox{\tiny nuc}}-g_\sigma\sigma$ and the degeneracy factor is $\gamma=2$ for asymmetric matter. The self-consistency of the model imposes to $M^*$ the condition of 
\begin{eqnarray}
M^*- M_{\mbox{\tiny nuc}} + \frac{g_\sigma^2}{m_\sigma^2}({\rho_s}_p + {\rho_s}_n) - \frac{A}{m_\sigma^2}\sigma^2 - \frac{B}{m_\sigma^2}\sigma^3 = 0,\quad 
\end{eqnarray}
with
\begin{eqnarray}
{\rho_s}_{p,n} &=& 
\frac{\gamma M^*}{2\pi^2}\int_0^{{k_F}_{p,n}}
\frac{k^2dk}{(k^2+M^{*2})^{1/2}}.
\label{rhospn}
\end{eqnarray}
The (field) equations for $\omega_0$ and $\bar{\rho}_{0(3)}$ are 
\begin{align}
m_\omega^2\omega_0 &= g_\omega\rho - Cg_\omega(g_\omega \omega_0)^3 
- g_\sigma g_\omega^2\sigma\omega_0(2\alpha_1+\alpha'_1g_\sigma\sigma)
\nonumber\\
&- \alpha'_3g_\omega^2 g_\rho^2\bar{\rho}_{0(3)}^2\omega_0
\end{align}
and
\begin{align}
m_\rho^2\bar{\rho}_{0(3)} &= \frac{g_\rho}{2}\rho_3 
-g_\sigma g_\rho^2\sigma\bar{\rho}_{0(3)}(2\alpha_2+\alpha'_2g_\sigma\sigma)
\nonumber\\
&-\alpha'_3g_\omega^2 g_\rho^2\bar{\rho}_{0(3)}\omega_0^2, 
\label{fieldrho}
\end{align}
with $\rho=\rho_p+\rho_n$, and $\rho_3=\rho_p-\rho_n=(2y-1)\rho$. The proton fraction of the system is $y=\rho_p/\rho$ and the proton/neutron densities are given by $\rho_{p,n}=\gamma{k_F^3}_{p,n}/(6\pi^2)$.

\section{Crustal properties of neutron stars (results)}
\label{crustal}

In order to correctly determine crustal properties of a neutron star, one needs, in principle, to have a complete description of its crust. In other words, the equation of state (EoS) for the inner and outer layers related to this part of the neutron star should be constructed. However, the description of the inner crust is a hard task since it is composed of different elements, namely, neutron-rich nuclei, nuclear clusters in the so-called pasta phase, and other. Furthermore, it is not trivial to construct a unified EoS that simultaneously describes the core and the crust of a neutron star. In particular, \cite{gognyic} constructed a unified EoS for the inner crust and the core computed with the D1M* Gogny force. As an alternative to this development, a treatment usually adopted to compute the EoS for the entire neutron star is based on a piecewise structure. In such an approach, the EoS is divided into 3 parts. One of them describes the outer crust, estimated to exist up to densities around $10^{11}$~g/cm$^3$~\cite{BPS-1971a,pieka}. For this part, the EoS developed by~\cite{BPS-1971a} is often used. For the core, hadronic relativistic and nonrelativistic models in which nucleons are the degrees of freedom are used from a density given by $\rho_t$, namely, the density related to the core-crust transition, to a region in which density reaches several times the saturation density. Finally, the inner crust (IC), region in between core and outer crust (OC), is widely treated through an effective EoS given by $P(\mathcal{E})=A+B\mathcal{E}^{4/3}$, with the constants $A$ and $B$ found by imposing two matchings, namely, OC-IC and IC-core.

Another approach proposed by~\cite{AA119-599} avoids a description for the crust by using that its mass ($M_{\rm crust}$) is negligible in comparison to the total neutron star mass~($M$). It leads (see Appendix) to expressions for $M_{\rm crust}$ and the crustal radius ($R_{\rm crust}$), namely,
\begin{eqnarray}
M_{\rm crust} = \frac{4\pi P_t R^4_{\rm core}}{GM_{\rm core}}\left(1-\frac{2GM_{\rm core}}{R_{\rm core}c^2}\right),
\label{mcrust}
\end{eqnarray}
and
\begin{eqnarray}
 R_{\rm crust} = \phi R_{\rm core} \left[\frac{1-R_{\rm s}/R_{\rm core}}{1-\phi\left(1-R_{\rm s}/R_{\rm core}\right)}\right],
\label{rcrust}
\end{eqnarray}
with $R_{\rm s} = 2GM/c^2$ and $\phi = [(\mu_t/\mu_0)^2 - 1]R_{\rm core}/R_{\rm s}$. Total mass and radius of the star are calculated as $M = M_{\rm core} +  M_{\rm crust}$ and $R = R_{\rm core} +  R_{\rm crust}$. $M_{\rm core}$ and $R_{\rm core}$ are mass and radius, respectively, determined by EoS of the core. $\mu_0 = \mu(P=0)$ is the chemical potential at the surface of the neutron star. Here, in this work, we consider $\mu_0 = 930.4$~MeV~\cite{heansel-book,PRC99-015803}.
In the equations presented above, $P_t$ and $\mu_t$ are quantities related to the core-crust transition. They are obtained here through the thermodynamical method described, for instance, by \cite{gogny7}, \cite{mdi4} and \cite{PRC100-015806}. This procedure is based on the search of mechanical and chemical stability region in stellar matter, in our case composed by neutron, protons, electrons and muons with charge neutrality and $\beta$-equilibrium conditions implemented, namely, $\mu_n - \mu_p = \mu_e$ and $\rho_p - \rho_e  = \rho_\mu$, where $\mu_{p,n}=\partial\epsilon_{\mbox{\tiny had}}/\partial\rho_{p,n}$, $\mu_e=(3\pi^2\rho_e)^{1/3}$, $\rho_\mu=[(\mu_\mu^2 - m_\mu^2)^{3/2}]/(3\pi^2)$, and $\mu_\mu=\mu_e$, for $m_\mu=105.7$~MeV and massless electrons. Total energy density and pressure of stellar matter are given by
\begin{eqnarray}
\mathcal{E} = \epsilon_{\mbox{\tiny had}} + \frac{\mu_e^4}{4\pi^2}
+ \frac{1}{\pi^2}\int_0^{\sqrt{\mu_\mu^2-m^2_\mu}}dk\,k^2(k^2+m_\mu^2)^{1/2},\qquad
\label{totaled}
\end{eqnarray}
and
\begin{eqnarray} 
P = p_{\mbox{\tiny had}} + \frac{\mu_e^4}{12\pi^2}
+ \frac{1}{3\pi^2}\int_0^{\sqrt{\mu_\mu^2-m^2_\mu}}\frac{dk\,k^4}{(k^2+m_\mu^2)^{1/2 }},\qquad\,
\label{totalp}
\end{eqnarray}
respectively.

\subsection{Selected parametrizations}
\label{selected}

\begin{figure}
\centering
\includegraphics[scale=0.34]{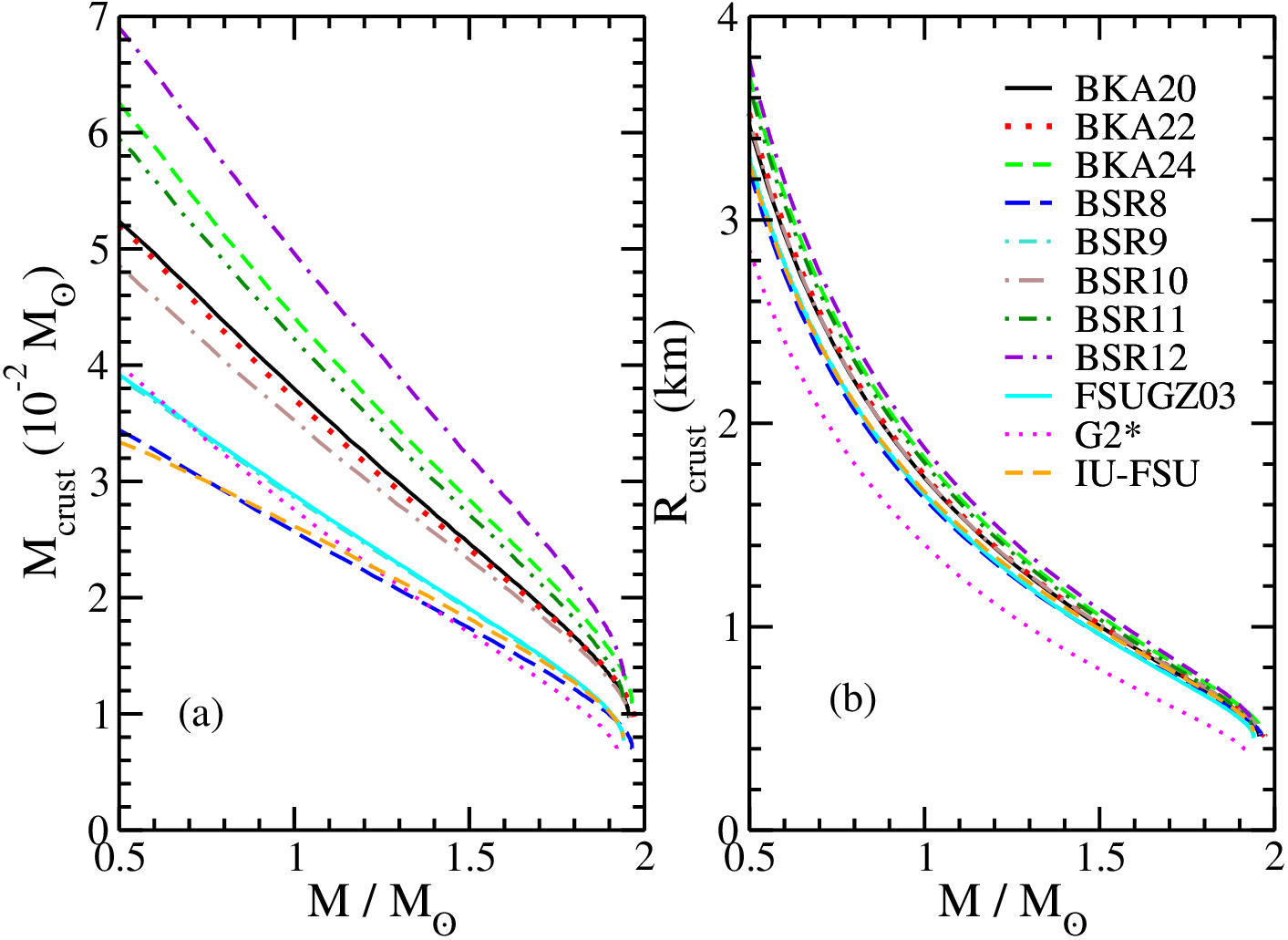}
\caption{(a) Mass, in units of $10^{-2}M_\odot$, and (b) radius of the crust as a function of $M/M_\odot$ for the RMF parametrizations of Table~\ref{tabcc}.}
\label{mrcrust}
\end{figure}

In Ref.~\cite{PRC90-055203}, 263 parametrizations of the RMF model described in previous section were tested against a set of constraints related to symmetric nuclear matter (SNM), pure neutron matter (PNM), symmetry energy and its slope. Such constraints include limits in pressure density dependence in SNM, ranges for incompressibility at the saturation density in SNM, boundaries in energy per particle density dependence in PNM, ranges for symmetry energy and its slope both at $\rho=\rho_0$, among other ones. The result of this analysis is that 35 parametrizations simultaneously satisfy constraints. Later on, they were also studied in the stellar matter regime in Ref.~\cite{PRC93-025806}. It was verified that neutron stars masses around two solar masses are obtained by the following parametrizations: BKA20~\cite{PRC81-034323}, BKA22~\cite{PRC81-034323}, BKA24~\cite{PRC81-034323}, BSR8~\cite{PRC76-045801}, BSR9~\cite{PRC76-045801}, BSR10 \cite{PRC76-045801}, BSR11~\cite{PRC76-045801}, BSR12~\cite{PRC76-045801}, FSUGZ03~\cite{PRC74-034323}, G2*~\cite{PRC74-045806}, \mbox{IU-FSU} \cite{PRC82-055803}. Their prediction for the mainly thermodynamic core-crust transition quantities are presented in Table~\ref{tabcc}.
\begin{table}[!htb]
\centering
\caption{Core-crust transition values for $\rho_t$ (fm$^{-3}$), $P_t$ (MeV/fm$^3$), $\mathcal{E}_t$ (MeV/fm$^3$), $y_t$ and $\rho_t/\rho_0$ for the parametrizations used in this work.}
\label{tabcc}       
\begin{tabular}{lccccc}
\hline\noalign{\smallskip}
Model & $\rho_t$    & $P_t$       & $\mathcal{E}_t$ & $y_t$ & $\rho_t/\rho_0$ \\
\noalign{\smallskip}\hline\noalign{\smallskip}
BKA20   & 0.070 & 0.345 & 66.613 & 0.026 & 0.483 \\ 
BKA22   & 0.066 & 0.330 & 62.865 & 0.026 & 0.452 \\
BKA24   & 0.066 & 0.381 & 62.525 & 0.026 & 0.450 \\
BSR8    & 0.073 & 0.256 & 68.903 & 0.032 & 0.495 \\
BSR9    & 0.072 & 0.283 & 67.842 & 0.031 & 0.487 \\
BSR10   & 0.070 & 0.334 & 65.921 & 0.030 & 0.472 \\
BSR11   & 0.068 & 0.383 & 64.592 & 0.028 & 0.465 \\
BSR12   & 0.073 & 0.494 & 68.890 & 0.030 & 0.494 \\
FSUGZ03 & 0.072 & 0.284 & 67.815 & 0.030 & 0.486 \\
G2*     & 0.069 & 0.284 & 65.009 & 0.021 & 0.448 \\
IU-FSU  & 0.091 & 0.320 & 85.969 & 0.044 & 0.586 \\
\noalign{\smallskip}\hline
\end{tabular}
\end{table}
Since $\rho_t$, $P_t$ and $\mu_t=(\mathcal{E}_t+P_t)/\rho_t$ are determined, it is possible to compute $M_{\rm crust}$ and $R_{\rm crust}$ if $M_{\rm core}$ and $R_{\rm core}$ are known. These last quantities are calculated through the Tolman–Oppenheimer–Volkoff (TOV) equations (\cite{tov39,tov39a}) given by (units in which $G=c=1$)
\begin{eqnarray}
\dfrac{dP(r)}{dr} &=& -\dfrac{\left[\mathcal{E}(r) + P(r)\right]\left[m(r) 
+ 4\pi r^3P(r)\right]}{r^2\left[1 - \dfrac{2m(r)}{r}\right]}
\label{tov1}
\end{eqnarray} 
and
\begin{eqnarray}
\dfrac{dm(r)}{dr} &=& 4\pi r^2 \mathcal{E}(r),
\label{tov2}
\end{eqnarray}
with solutions constrained to the following two conditions at the neutron star 
center: $P(0) = P_c$ (central pressure), and $m(0) = 0$ (central mass). In this case, energy density and pressure are those exclusively presented in Eqs.~(\ref{totaled}) and ~(\ref{totalp}) starting from the transition density $\rho_t$, i.e., with no crust EoS added. Thus, the mass-radius profile obtained from the TOV equations is only related to the neutron star core in which $m(R_{\rm core})=M_{\rm core}$. Such a procedure allows to use Eqs.~(\ref{mcrust}) and~(\ref{rcrust}), with results depicted in Fig.~\ref{mrcrust}. 
\begin{figure}
\centering
\includegraphics[scale=0.32]{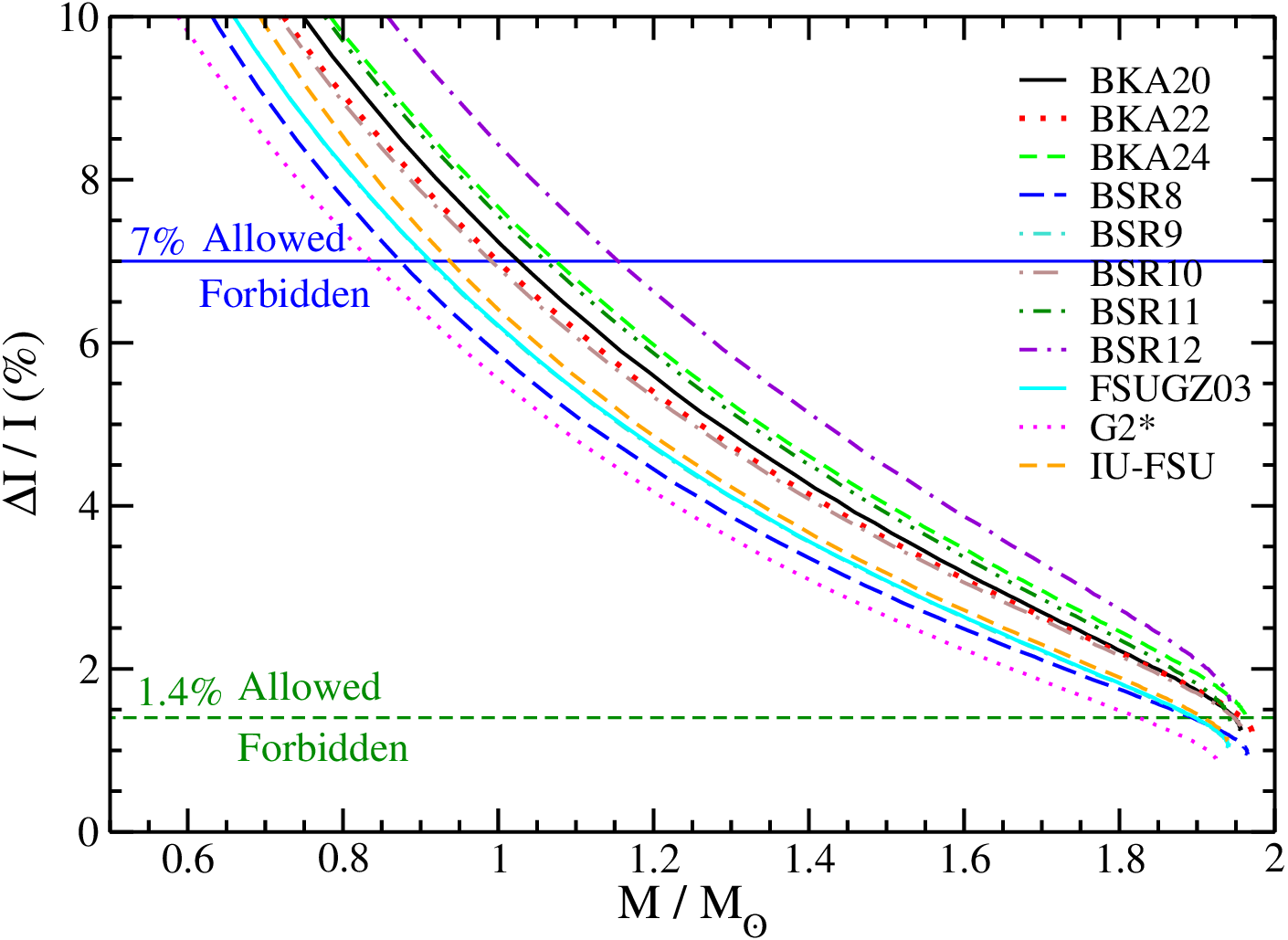}
\caption{Crustal fraction of the moment of inertia versus $M/M_\odot$ for the RMF parametrizations of Table~\ref{tabcc}. Horizontal lines: boundaries related to the $\Delta I/I \geqslant 1.4\%$ and $\Delta I/I \geqslant 7\%$ constraints.}
\label{deltai}
\end{figure}
From Fig.~\ref{mrcrust}{\color{blue}a} we verify that only a very small part of the neutron star mass is located at its crust. As an example, for the $M=1.4M_\odot$, we obtain a range of around $1.4\%$~(G2*) to $2.6\%$~(BSR12) of the total mass enclosed by the crust, predicted by the RMF parametrizations studied here. This finding is in agreement with other studies as the reader can verify in Refs.~\cite{PRC100-015806,pieka10,pieka14}, for instance. The situation changes when radius is analyzed. For the same parametrizations, Fig.~\ref{mrcrust}{\color{blue}b}, we notice that the crust contribution for this quantity is more significant since the range of the total radius accounted by the crust is given by $7.1\%$~(G2*) to $9.2\%$~(BSR12). This is a feature also observed in other parametrizations. In Ref.~\cite{pieka14}, \mbox{TFcmax} model, for example, predicts a crust encompassing about $17\%$ of the $M=1.4M_\odot$ star radius.

Now we calculate the ratio between the crustal fraction of the moment of inertia and the total moment of inertia, $\Delta I/I$, for the RMF parametrizations selected before. With the aim of obtaining such a quantity still not using an specific EoS for the neutron star crust, or neither constructing unified EoS's for the entire description of the compact object, we use the approximation for $\Delta I/I$ given in Refs.~\cite{Lattimer-2001,Lattimer-2007,Madhuri-2017,PRC100-015806,PR121-333} and written as
\begin{eqnarray}
\frac{\Delta I}{I} &=& \frac{28\pi P_tR^3}{3Mc^2}\frac{(1-1.67\xi-0.6\xi^2)}{\xi} 
\nonumber \\
&\times& \left[1+\frac{2\, P_t}{\rho_t \, m_b\, c^2}\frac{(1+5\xi-14\xi^2)}{\xi^2}\right]^{-1},
\label{deli}
\end{eqnarray}
with $\xi = GM/(Rc^2)$ being the dimensionless compactness, and $m_b = 930$~MeV/c$^2$ the mass of $^{12}\rm C/12$ or $^{56}\rm Fe/56$~\cite{Lattimer-2000,Lattimer-2001}.

\begin{figure*}
\centering
\includegraphics[scale=0.55]{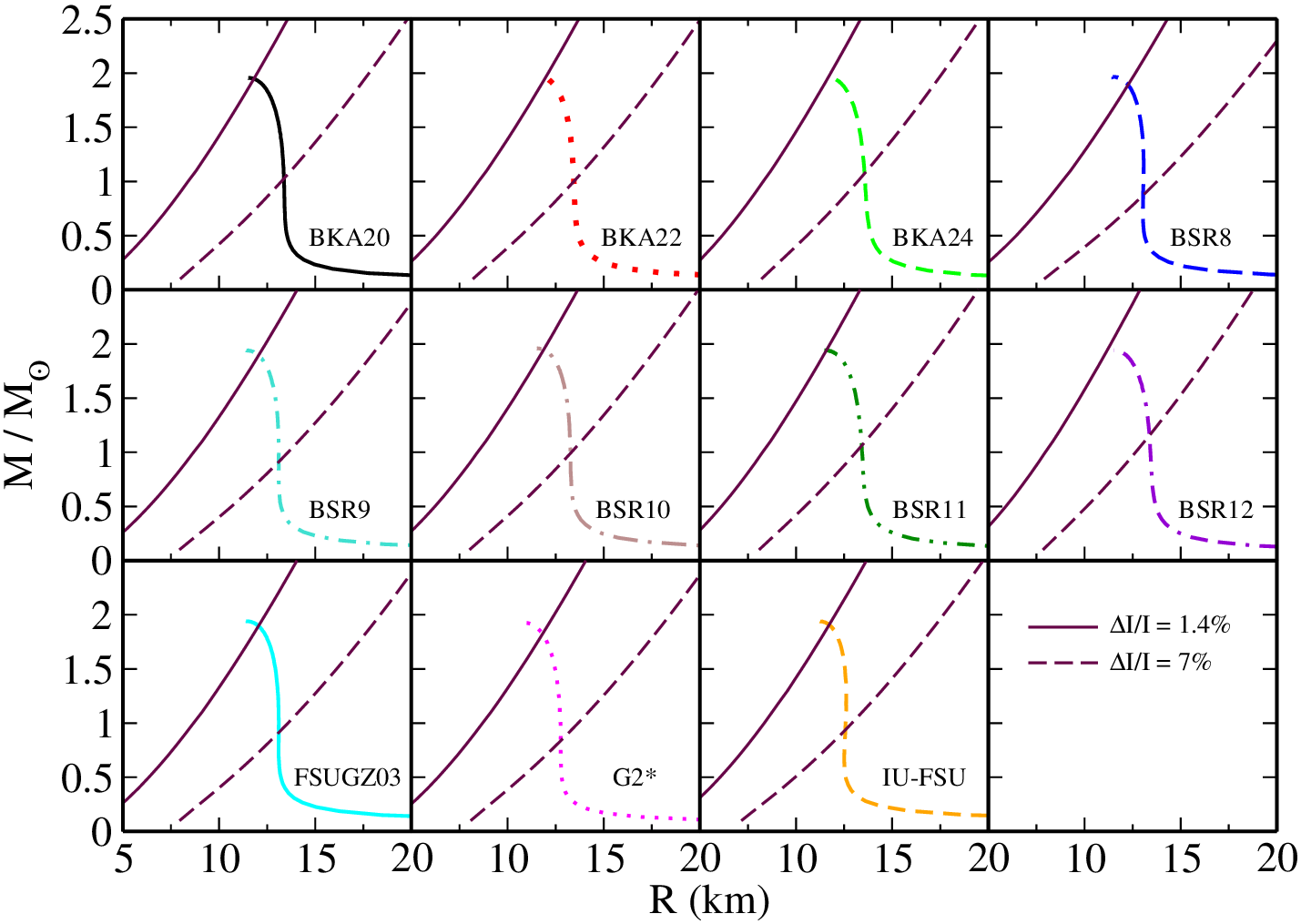}
\caption{Mass-radius diagrams for the RMF parametrizations, along with curves (see text) in which $\Delta I/I = 1.4\%$ (full) and~$7\%$~(dashed).}
\label{mr}
\end{figure*}

Total mass and radius used as inputs in Eq.~(\ref{deli}) are obtained from $M = M_{\rm core} +  M_{\rm crust}$ and $R = R_{\rm core} +  R_{\rm crust}$, along with quantities given in Eqs.~(\ref{mcrust}) and~(\ref{rcrust}). The results are shown in Fig.~\ref{deltai}. In this figure we also display the lower limits of $\Delta I/I \geqslant 1.4\%$ and $\Delta I/I \geqslant 7\%$, estimated for the Vela pulsar that explain the observed glitching mechanism. The latter (former) limit refers to the analysis with (without) entrainment effects included in the estimation of $\Delta I/I$~\cite{PRL109-241103,PRL110-011101}. We verify that all parametrizations predict $\Delta I/I \geqslant 1.4\%$ for a range of $M\leqslant 1.82M_\odot$, with the G2* model defining this boundary. The ``forbidden'' region is very small in comparison to the ``allowed'' one. The same does not apply if we take into account the $\Delta I/I \geqslant 7\%$ constraint. In this case, one sees that allowed region defined by parametrizations is those in which $M\leqslant 1.16M_\odot$. The upper limit is determined by BSR12 model. Another way to identify the regions defined by the $\Delta I/I$ constraints is directly from the mass-radius profiles. Such curves are constructed using the method proposed by Zdunik et al~\cite{AA119-599}, i.e., without a specific description for the neutron star crust. The results are displayed in Fig.~\ref{mr}.
The boundary curves related to $\Delta I/I=1.4\%$ constraint were built as follows. $\Delta I/I$ is a function of $M$ and $R$, since $P_t$ is known for the models. Therefore, we fix $\Delta I/I$ in $1.4\%$ and running $M$. For each $M$, we invert Eq.~(\ref{deli}) and find the respective value for $R$. Then, the curve $M\times R$ is determined with the condition $\Delta I/I$ in $1.4\%$ satisfied. The same procedure is adopted when $\Delta I/I=7\%$ constraint is taken into account. The allowed (forbidden) regions are those at the right (left) of the contour curves. Notice that masses and radii are much more restricted for the case of $\Delta I/I \geqslant 7\%$. For this reason, we present in Table~\ref{tabfit} an estimation for the coefficients of the approximation given by 
\begin{eqnarray}
R \geqslant c + b(M/M_\odot) - a(M/M_\odot)^2
\label{rvela}
\end{eqnarray}
for the Vela pulsar radius, by taking into account only the $\Delta I/I \geqslant 1.4\%$ case. 
\begin{table}
\centering
\caption{Estimation for the Vela pulsar radius as a function of $M$ based on the $\Delta I/I \geqslant 1.4\%$ constraint.}
\label{tab:fit-radius}       
\begin{tabular}{lccc}
\hline\noalign{\smallskip}
Parametrization &   $c$ (km) & $b$ (km) & $a$ (km) \\
\noalign{\smallskip}\hline\noalign{\smallskip}
BKA20   & $3.922$ & $4.844$ & $0.401$\\
BKA22   & $4.004$ & $4.871$ & $0.403$\\
BKA24   & $3.967$ & $4.678$ & $0.374$\\
BSR8    & $3.999$ & $5.261$ & $0.462$\\
BSR9    & $3.974$ & $5.117$ & $0.441$\\
BSR10   & $3.944$ & $4.878$ & $0.405$\\
BSR11   & $3.926$ & $4.687$ & $0.377$\\
BSR12   & $3.797$ & $4.491$ & $0.349$\\
FSUGZ03 & $3.972$ & $5.111$ & $0.440$\\
G2*     & $4.016$ & $5.093$ & $0.436$\\
IU-FSU  & $3.699$ & $5.048$ & $0.433$\\
\noalign{\smallskip}\hline
\label{tabfit}
\end{tabular}
\end{table}
This expression is based on the analysis of the results obtained in Fig.~\ref{mr} with entrainment effects ignored.

\subsection{Bulk parameters effect}

Now we turn our attention to analyze how nuclear matter bulk parameters affects the crustal properties of neutron stars. In order to do that, we start by defining a initial parametrization of the RMF model described in previous section. Six coupling constants of the model, namely, $g_\sigma$, $g_\omega$, $g_\rho$, $A$, $B$, and $\alpha'_3$ are determined in order to reproduce six bulk parameters. They are: $\rho_0=0.15$~fm$^{-3}$, $B_0=-16.0$~MeV (binding energy), $m^*\equiv M^*_0/M_{\mbox{\tiny nuc}}=0.60$, $K_0=230$~MeV (incompressibility at $\rho=\rho_0$), $J=31.6$~MeV (symmetry energy at $\rho=\rho_0$) and $L_0=58.9$~MeV (symmetry energy slope at $\rho=\rho_0$). Here, one has $B_0=E(\rho_0) - M$, 
$M_0^*=M^*(\rho_0)$, $K_0=9(\partial p/\partial\rho)_{\rho_0}$, $J=\mathcal{S}(\rho_0)$, and 
$L_0=3\rho_0(\partial\mathcal{S}/\partial\rho)_{\rho_0}$, with 
$\mathcal{S}(\rho)=(1/8)(\partial^{2}E/\partial y^2)|_{y=1/2}$ and $E(\rho)=\epsilon_{\mbox{\tiny had}}/\rho$. We also use $m_\omega = 782.5$~MeV, $m_\rho = 763$~MeV and $m_\sigma=500$~MeV in Eqs.~(\ref{denerg})-(\ref{fieldrho}). The coupling constant $C$ is chosen to be equal to $0.005$. This value ensures that the model predicts neutron stars with masses around $2M_\odot$~\cite{lucas}. The remaining free parameters, namely, $\alpha_1$, $\alpha'_1$, $\alpha_2$, and $\alpha'_2$ are taken as vanishing for the sake of simplicity.
We remark that the choice of such numbers for the bulk parameters is based on different theoretical and 
experimental studies. Saturation density and binding energy are well established closely around the values of $0.15$~fm$^{-3}$ and $-16.0$~MeV, respectively, in many different hadronic models~\cite{PRC90-055203,stone,had1,apj}. Regarding the symmetry energy and its slope, authors of Ref.~\cite{plb2013} point out to the ranges of $J=31.6\pm 2.66$~MeV and $L_0=58.9\pm 16$~MeV based on analyses of different terrestrial nuclear experiments and astrophysical observations. Similar numbers, namely, $J= 31.7 \pm 3.2$~MeV and $L_0=58.7\pm 28.1$~MeV, were also found in Ref.~\cite{jlrange2}. With regard to the incompressibility, $K_0=230$~MeV is consistent with the range of $220\,\mbox{MeV}\leqslant K_0 \leqslant 260~\mbox{MeV}$ of Ref.~\cite{k4}, for instance. Lastly, the value $m^*=0.6$ is compatible with the limits of $0.58\leqslant m^* \leqslant 0.64$ obtained in Ref.~\cite{furns}. 

\begin{figure}
\centering
\includegraphics[scale=0.35]{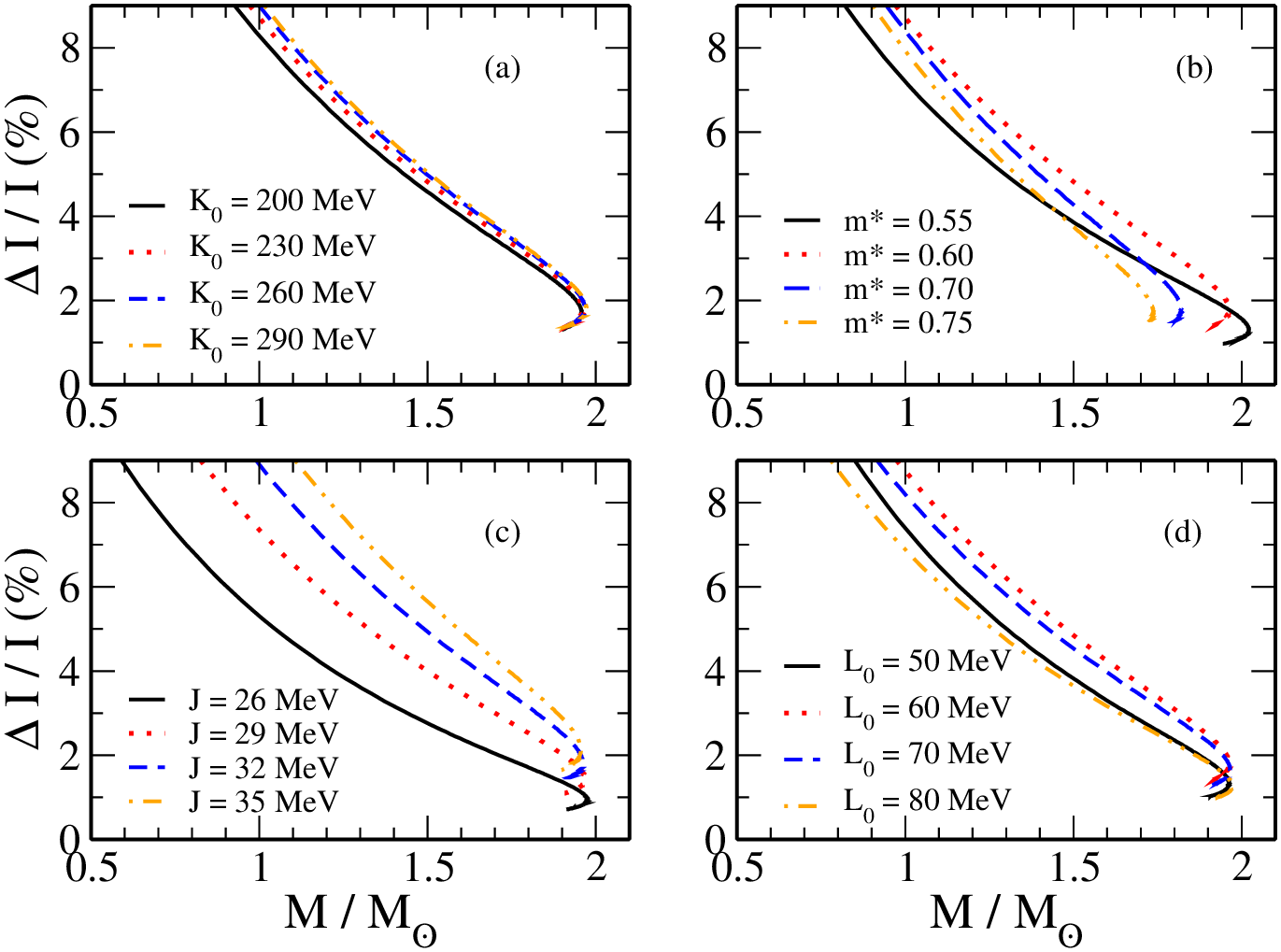}
\caption{Crustal fraction of the moment of inertia versus $M/M_\odot$ for different parametrizations of the RMF model obtained by varying (a)~$K_0$, (b)~$m^*$, (c)~$J$ and (d)~$L_0$.}
\label{deltai-corr}
\end{figure}

By starting with the initial model, we generate different parametrizations obtained from the independent variation of each bulk parameter. In this way, we ensure that the impact in the quantity analyzed is specifically due to the bulk parameter we are varying. This study was firstly performed for crust mass and radius, with results depicted in next Figs.~\ref{mcrust-corr} and ~\ref{rcrust-corr}.
\begin{figure}
\centering
\includegraphics[scale=0.35]{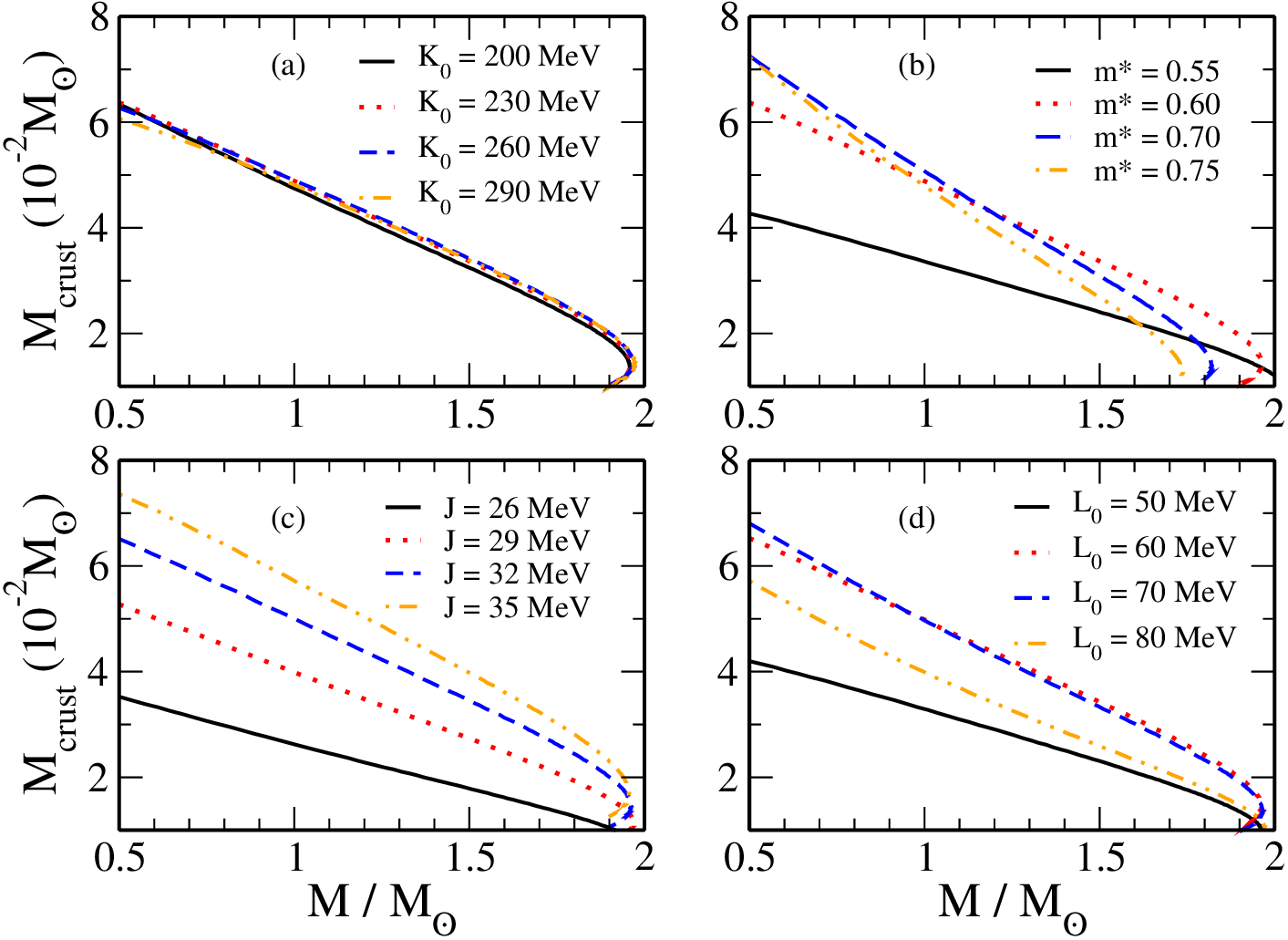}
\caption{$M_{\rm crust}$, in units of $10^{-2}M_\odot$, as a function of $M/M_\odot$ for different parametrizations of the RMF model obtained by varying (a)~$K_0$, (b)~$m^*$, (c)~$J$ and (d)~$L_0$.}
\label{mcrust-corr}
\end{figure}
\begin{figure}
\centering
\includegraphics[scale=0.35]{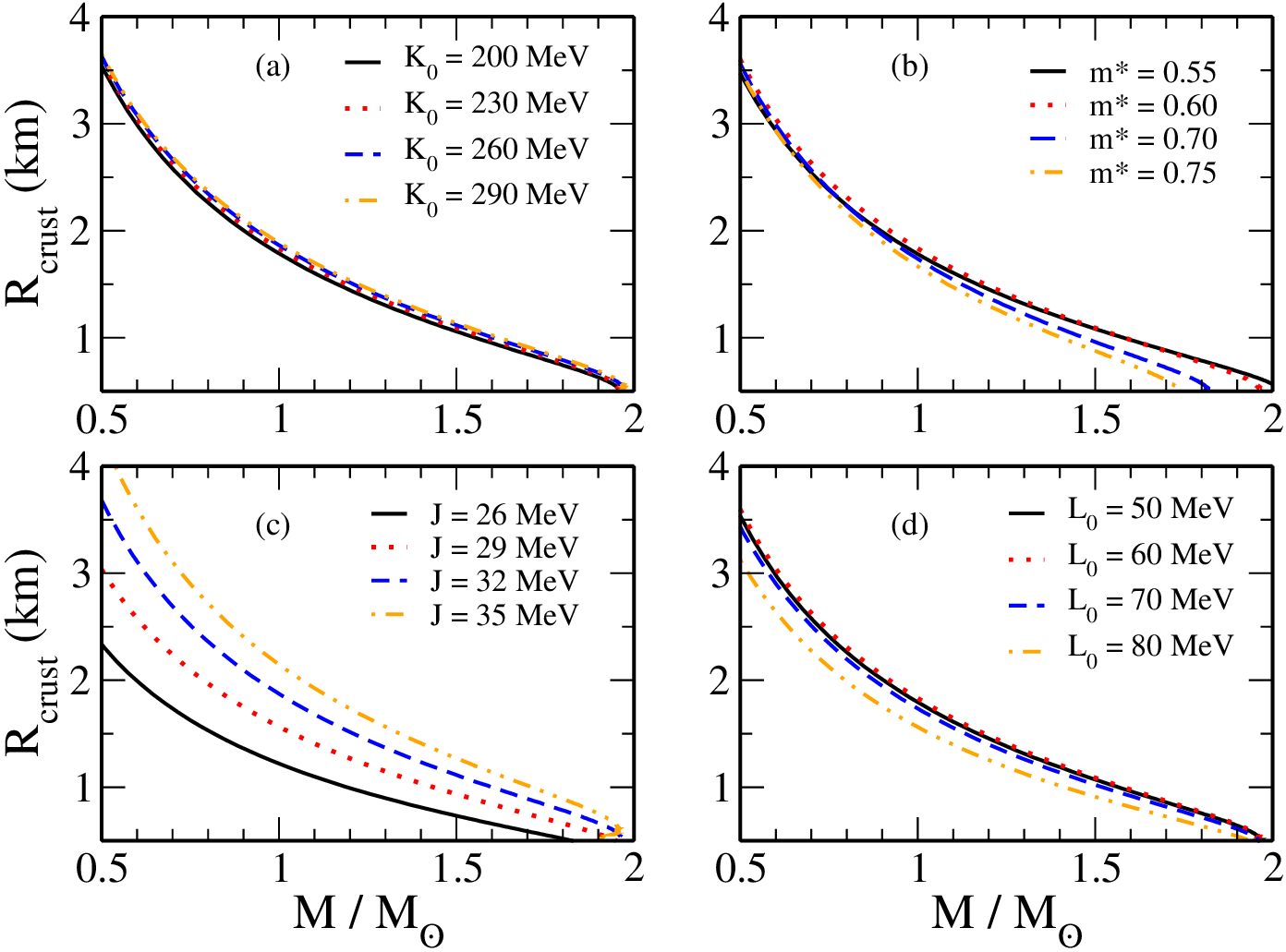}
\caption{$R_{\rm crust}$ as a function of $M/M_\odot$ for different parametrizations of the RMF model obtained by varying (a)~$K_0$, (b)~$m^*$, (c)~$J$ and (d)~$L_0$.}
\label{rcrust-corr}
\end{figure}
In Figs.~\ref{mcrust-corr}{\color{blue}a} and ~\ref{rcrust-corr}{\color{blue}a}, for instance, the four parametrizations present fixed values of $\rho_0=0.15$~fm$^{-3}$, $B_0=-16.0$~MeV, $m^*=0.60$, $J=31.6$~MeV and $L_0=58.9$~MeV, but different $K_0$ values, namely, $K_0=200$, $230$, $260$ and $290$~MeV. The same procedure was adopted for the other isoscalar and isovector bulk parameters in panels (b), (c) and (d) of the figures. From Fig~\ref{mcrust-corr}, one can see that $K_0$ almost produce no impact in the mass of the crust (panel a), unlike $J$, that exerts the higher effect (panel c). $M_{\rm crust}$ is also susceptible to variations of effective mass, $m^*$ (panel b), and slope of the symmetry energy $L_0$ (panel d). In the case of $R_{\rm crust}$, on the other hand, the variation of bulk parameters does not to play a significant role, except for the symmetry energy, $J$, as one can see in Fig.~\ref{rcrust-corr}{\color{blue}c}. Regarding the effect of $J$ on $M_{\rm crust}$ and $R_{\rm crust}$, it is worth to notice that a pattern is observed for all range of $M$ analyzed, namely, increasing of $M_{\rm crust}$ and $R_{\rm crust}$ as $J$ increases, see Figs.~\ref{mcrust-corr}{\color{blue}c} and~\ref{rcrust-corr}{\color{blue}c}. We also investigate how the bulk parameters variation affects the ratio between the crustal fraction of the moment of inertia and the total moment of inertia $\Delta I/I$. This is a quantity that directly depends on $M_{\rm crust}$ and $R_{\rm crust}$, according to Eqs.~(\ref{deli}), thus the results presented before directly impact $\Delta I/I$. The outcomes are given in Fig.~\ref{deltai-corr}. 
As we verify, the symmetry energy is the bulk parameter that produces higher variations in the crustal moment of inertia as also observed in the case of $M_{\rm crust}$ and $R_{\rm crust}$. On the other hand, $K_0$ is found, once again, to be the quantity that produces almost no changes. $L_0$ and $m^*$ have some effect on $\Delta I/I$ by impacting it in a strength between the ones exhibited by $K_0$ and $J$.

\begin{figure}
\centering
\includegraphics[scale=0.35]{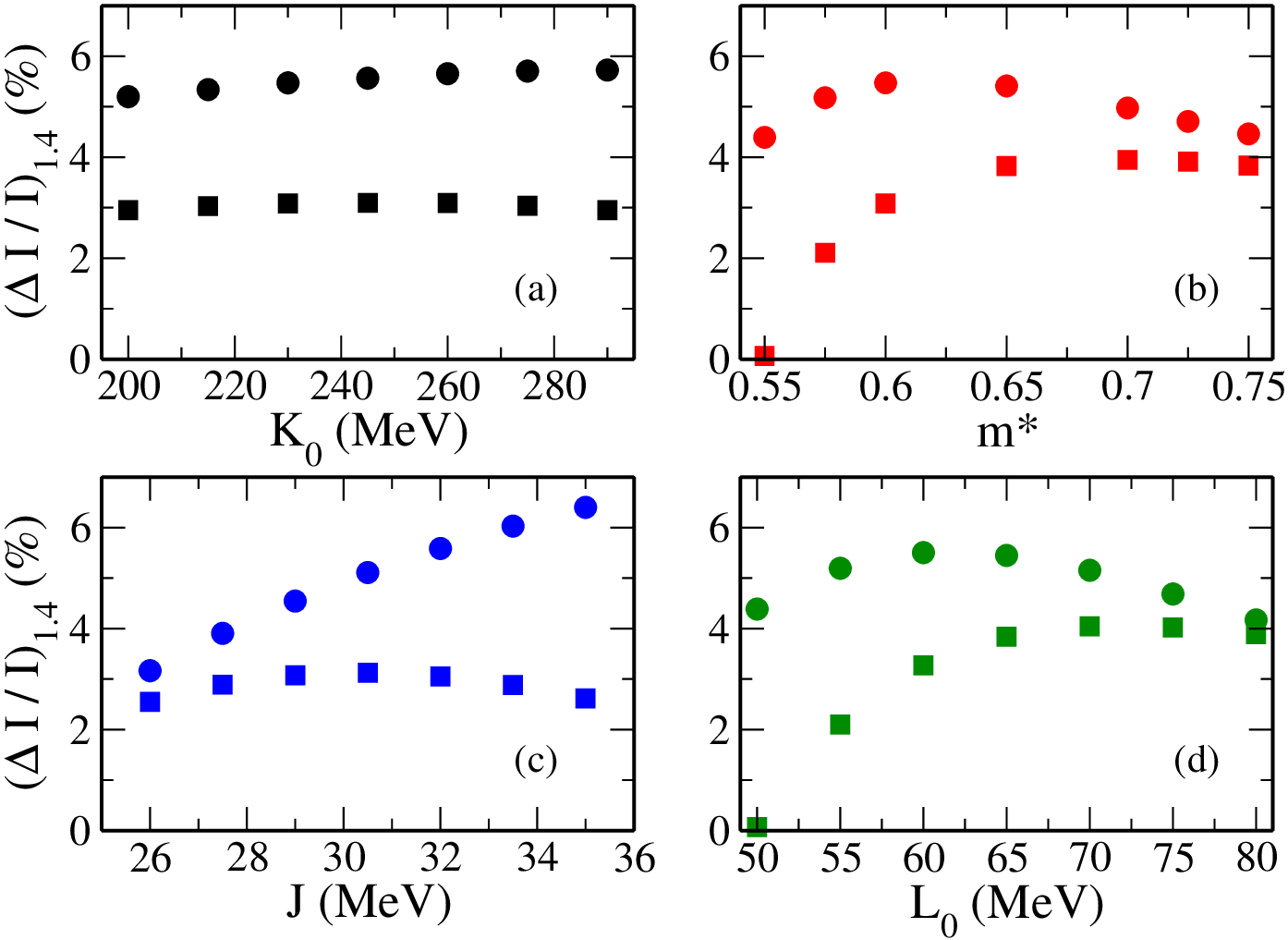}
\caption{Crustal fraction of the moment of inertia for $M=1.4M_\odot$ as a function of (a)~$K_0$, (b)~$m^*$, (c)~$J$ and (d)~$L_0$ for cases in which $\alpha_1=\alpha'_1=\alpha_2=\alpha'_2=0$ (circles) and $\alpha_1= 0.2746$~MeV, $\alpha_2 = 5.6995$~MeV, $\alpha'_1 =2.6680\times10^{-4}$, $\alpha'_2 = 5.5594\times10^{-3}$ (squares).}
\label{deltai14}
\end{figure}

\begin{figure}
\centering
\includegraphics[scale=0.32]{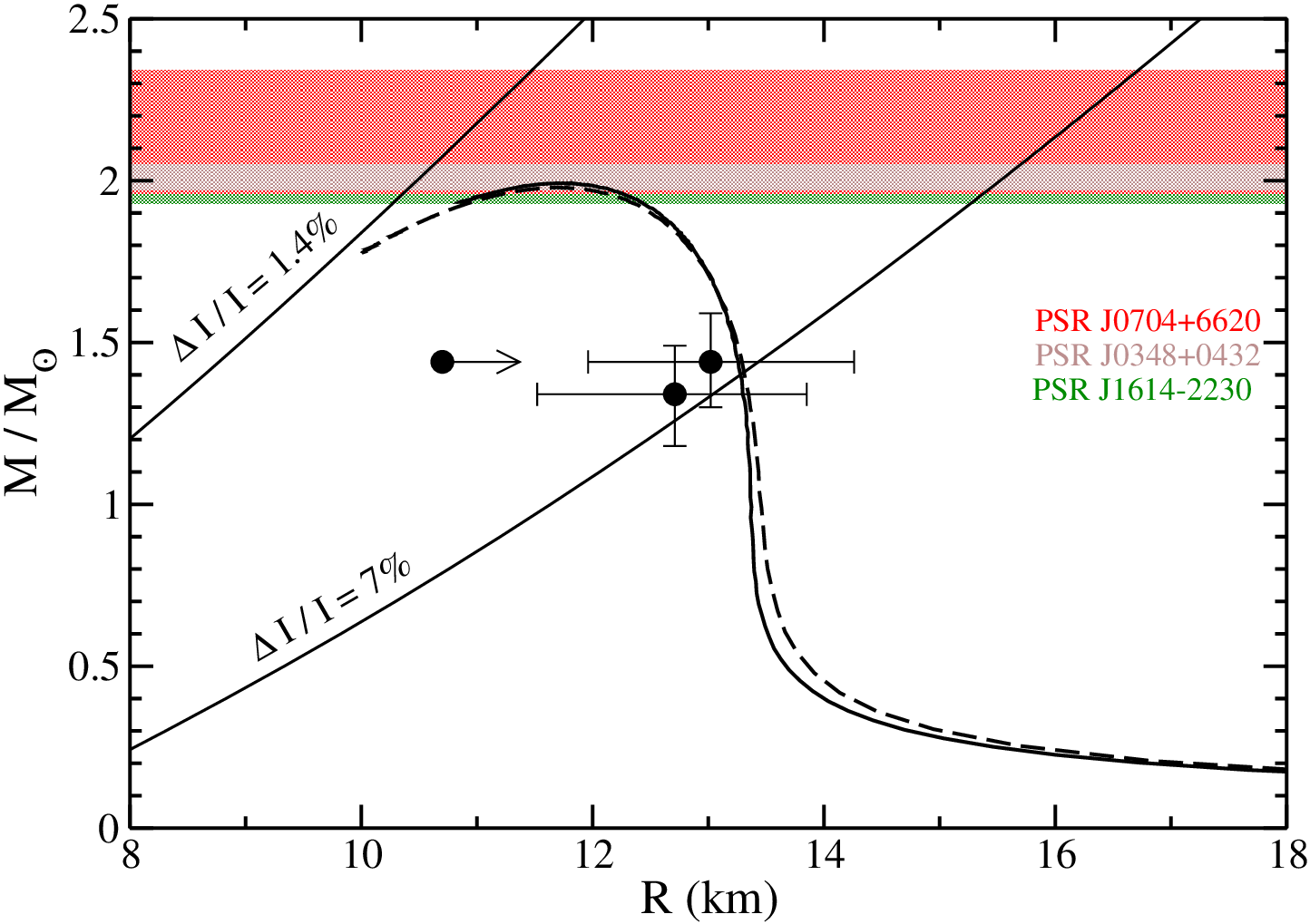}
\caption{Mass-radius profile for the RMF parametrization constructed to satisfy the $\Delta I/I \geqslant 7\%$ constraint at $M=1.4M_\odot$. Bands extracted from Refs.~\cite{Demorest,Antoniadis,Cromartie}. Circles with error bars: NICER data~(\cite{Col,l21,l25}). Full lines: curves related to the case of $\alpha_1=\alpha'_1=\alpha_2=\alpha'_2=0$. Dashed line: case of $\alpha_1= 0.2746$~MeV, $\alpha_2 = 5.6995$~MeV, $\alpha'_1 =2.6680\times10^{-4}$, $\alpha'_2 = 5.5594\times10^{-3}$}.
\label{mr-new}
\end{figure}

Since we have information regarding how $K_0$, $J$, $m^*$ and $L_0$ impact on $\Delta I/I$ in a neutron star mass range up to around $2M_\odot$, it is natural to search for a particular parametrizations that satisfies the constraint of $\Delta I/I \geqslant 7\%$, the condition that explains the glitching mechanism when entrainment effect are taken into account, for masses around $M=1.4M_\odot$. This particular neutron star mass value was used in Ref.~\cite{pavlov} to fit data from the softer component of the X-ray spectrum, detected by the Chandra X-Ray Observatory, for the Vela pulsar. As we previously discussed, the selected RMF parametrizations studied in Sec.~\ref{selected} attain $\Delta I/I \geqslant 7\%$ for a range of $M\leqslant 1.16M_\odot$, with the upper limit given by the BSR12 model. If we want to define a particular parametrization with this purpose, it is useful to analyze how $\Delta I/I$ for $M=1.4M_\odot$ depends on the bulk parameters. These findings are represented by the circles in Fig.~\ref{deltai14}.
Therefore, we were able to generate a RMF parametrization presenting the following quantities: $\rho_0=0.15$~fm$^{-3}$, $B_0=-16.0$~MeV, $m^*=0.575$, $K_0=260$~MeV, $J=35$~MeV, and $L_0=70$~MeV. With regard to the value of $L_0$, it is low but still inside the range recently proposed in Ref.~\cite{piekaprex2}, namely, $69\leqslant L_0\leqslant 143$~MeV, determined from an analysis of the results provided by the PREX-II collaboration concerning the neutron skin thickness in $^{208}\rm Pb$~\cite{prex2}. However, it is in full agreement with another estimation predicting $42\leqslant L_0\leqslant 117$~MeV, obtained from measurements of the spectra of charged pions~\cite{pions}. The mass-radius profile of this model, constructed through the Zdunik method, is displayed in Fig.~\ref{mr-new} together with curves of $\Delta I/I = 1.4\%$ and~$\Delta I/I = 7\%$.
From the figure, we also verify that the model is in agreement with observational data from PSR J1614-2230, $M=1.97\pm 0.04M_{\odot}$~\cite{Demorest}, PSR J0348+0432, $M=2.01\pm 0.04M_{\odot}$~\cite{Antoniadis}, and the recent one related to the MSP J0740+6620 pulsar at $95.4\%$ credible level, namely, $M=2.14^{+0.20}_{-0.18} M_{\odot}$~\cite{Cromartie}. It is worth to notice that the model is also compatible with with recent data from the Neutron Star Interior Composition Explorer (NICER) mission, namely, $M=1.44^{+0.15}_{-0.14}M_{\odot}$ with $R=13.02^{+1.24}_{-1.06}$~km~\cite{Col}, $M=1.34^{+0.15}_{-0.16}M_{\odot}$ with $R=12.71^{+1.14}_{-1.19}$~km~\cite{l21}, and $R_{1.44}>10.7$~km~\cite{l25}. Finally, we use the agreement of the model with the $\Delta I/I \geqslant 7\%$ constraint to estimate the radius for the Vela pulsar with entrainment effects included. The result is the expression given by Eq.~(\ref{rvela}) with $a=0.39817$~km, $b=4.43579$~km, and $c=3.24293$~km.

For the sake of completeness, we also analyze a more generalized structure of the aforementioned RMF model by including in our study cases in which the mesons interactions $\rho-\sigma$ and $\omega-\sigma$ are also taken into account. This is done by defining particular values for the constants $\alpha_1$, $\alpha'_1$, $\alpha_2$, and $\alpha'_2$, previously set to zero. Specifically, we adopt the mean values of each quantity calculated from the parametrizations used in Sec.~\ref{selected}. The numbers are the following: $\alpha_1= 0.2746$~MeV, $\alpha_2 = 5.6995$~MeV, $\alpha'_1 =2.6680\times10^{-4}$, and $\alpha'_2 = 5.5594\times10^{-3}$. In Table~\ref{tabalpha}, we furnish some star properties calculated from the $16$ possible combinations constructed by making vanishing $(-)$ or not ($\checkmark$) that specific coupling constant.
\begin{table*}[!htb]
\centering
\caption{Some neutron star properties obtained from the RMF model with the interaction in which the corresponding nonvanishing strength is marked with the symbol $\checkmark$. For the entire set we use $\rho_0=0.15$~fm$^{-3}$, $B_0=-16.0$~MeV, $m^*=0.60$, $K_0=230$~MeV, $J=31.6$~MeV, and $L_0=58.9$~MeV.}
\begin{tabular}{cccccccccc}
\hline\noalign{\smallskip}
$\alpha_1$ &  $\alpha'_1$ & $\alpha_2$ & $\alpha'_2$ 
& $M_{\mbox{\tiny max}}$ ($M_\odot$)   & $R_{\mbox{\tiny max}}$ (km) 
& $M_{\mbox{\tiny crust}}$ ($M_\odot$) & $R_{\mbox{\tiny crust}}$ (km) 
& $(\Delta I/I)_{1.4}$ ($\%$)&  $I$~(MeV.km$^2$) \\
\noalign{\smallskip}\hline\noalign{\smallskip}
    $-$     &    $-$     &    $-$     &    $-$     & 1.963 & 11.457 & 3.677 & 1.211 & 5.471 & 80.693\\
$\checkmark$&$\checkmark$&$\checkmark$&$\checkmark$& 1.933 & 11.311 & 1.691 & 1.061 & 3.085 & 80.391\\
$\checkmark$&   $-$      &  $-$       &   $-$      & 1.939 & 11.376 & 3.521 & 1.190 & 5.278 & 79.819\\
    $-$     &$\checkmark$&  $-$       &    $-$     & 1.948 & 11.364 & 3.580 & 1.199 & 5.358 & 80.195\\
  $-$       &    $-$     &$\checkmark$&  $-$       & 1.968 & 11.469 & 1.902 & 1.092 & 3.405 & 81.548\\
   $-$      &    $-$     &    $-$     &$\checkmark$& 1.964 & 11.416 & 3.593 & 1.205 & 5.386 & 80.811\\
$\checkmark$&$\checkmark$&    $-$     &   $-$      & 1.926 & 11.313 & 3.432 & 1.179 & 5.175 & 79.355\\
$\checkmark$&    $-$     &$\checkmark$&    $-$     & 1.944 & 11.413 & 1.824 & 1.074 & 3.274 & 80.681\\
$\checkmark$&    $-$     &    $-$     &$\checkmark$& 1.941 & 11.364 & 3.445 & 1.185 & 5.200 & 79.947\\
    $-$     &$\checkmark$&$\checkmark$&    $-$     & 1.954 & 11.416 & 1.838 & 1.082 & 3.310 & 81.069\\
     $-$    &$\checkmark$&   $-$      &$\checkmark$& 1.950 & 11.371 & 3.505 & 1.195 & 5.284 & 80.344\\
     $-$    &    $-$     &$\checkmark$&$\checkmark$& 1.970 & 11.487 & 1.825 & 1.088 & 3.301 & 81.727\\
$\checkmark$&$\checkmark$&$\checkmark$&   $-$    & 1.930 & 11.381 & 1.764 & 1.065 & 3.186 & 80.232\\
$\checkmark$&$\checkmark$&  $-$       &$\checkmark$& 1.927 & 11.305 & 3.350 & 1.173 & 5.089 & 79.451\\
$\checkmark$&    $-$     &$\checkmark$&$\checkmark$& 1.946 & 11.437 & 1.747 & 1.069 & 3.170 & 80.828\\
      $-$   &$\checkmark$&$\checkmark$&$\checkmark$& 1.956 & 11.442 & 1.759 & 1.077 & 3.200 & 81.219\\
\noalign{\smallskip}\hline
\label{tabalpha}
\end{tabular}
\end{table*}

In the first line of this table we show results related to the case $\alpha_1=\alpha'_1=\alpha_2=\alpha'_2=0$ used as a guide to generate the circles in Fig.~\ref{deltai14}, and used to establish the parametrization whose mass-radius diagram is displayed as a full line in Fig.~\ref{mr-new}. Notice that this is the case that produces the higher value of $(\Delta I/I)_{1.4}$, around $5.5\%$. Any other combination of the $\rho-\sigma$ and $\omega-\sigma$ interactions presents a lower value for this quantity. In Fig.~\ref{deltai14}, for instance, we also plot curves generated by the case in which all constants are different from zero (squares in each panel). Notice that all of them are below the previous curves (circles). Despite these findings, we remark that it is still possible to find a parametrization that leads to $(\Delta I/I)_{1.4}=7\%$ with a particular configuration different from the $\alpha_1=\alpha'_1=\alpha_2=\alpha'_2=0$ case. As an example, the case in which the model exhibits 
$\alpha_1\neq 0$ and $\alpha'_1=\alpha_2=\alpha'_2=0$ can be used to construct a parametrization presenting $(\Delta I/I)_{1.4}=7.2\%$. Its bulk parameters are given by $\rho_0=0.15$~fm$^{-3}$, $B_0=-16.0$~MeV, $m^*=0.565$, $K_0=260$~MeV, $J=36$~MeV, and $L_0=74$~MeV. The mass-radius profile obtained from this new model is also shown in Fig.~\ref{mr-new}, see the dashed line. However, notice that this curve does not differ significantly from the previous parametrization determined by taking $\alpha_1=\alpha'_1=\alpha_2=\alpha'_2=0$.

\section{Summary and Concluding remarks}
\label{conclusions}

In this work we analyze the outcomes related to the crustal properties of neutron stars provided by a set of relativistic hadronic mean-field (RMF) model parametrizations consistent with different constraints coming from symmetric and asymmetric nuclear matter, pure neutron matter, and some astrophysical data related studied in Refs.~\cite{PRC90-055203,PRC93-025806}. We use the approach in which it is possible to calculate crust mass ($M_{\rm crust}$) and radius ($R_{\rm crust}$) without a specific treatment for this part of the neutron star~\cite{AA119-599}. Since this quantities are obtained, it is possible to determine analytically the ratio between the crustal fraction of the moment of inertia and the total moment of inertia, $\Delta I/I$ given in Eq.~(\ref{deli}), and verify which parametrizations satisfy the constraints of $\Delta I/I \geqslant 1.4\%$ and $\Delta I/I \geqslant 7\%$, found to be important to correctly explain the glitching mechanism observed in pulsars, such as the Vela one~\cite{pulsars}, with and without entrainment effects included. 

We found that all parametrizations studied predict $\Delta I/I \geqslant 1.4\%$ for a range of $M\leqslant 1.82M_\odot$. If we take into account the $\Delta I/I \geqslant 7\%$ constraint, our findings indicate a smaller range of $M\leqslant 1.16M_\odot$. One can identify the regions compatible with these boundaries in the mass-radius diagrams exhibited in Fig.~\ref{mr}. By taking into account the first restriction ($\Delta I/I \geqslant 1.4\%$), we were able to estimate the radius for the Vela pulsar as $R \geqslant c + b(M/M_\odot) - a(M/M_\odot)^2$, with coefficients predicted by the RMF parametrizations given in Table.~\ref{tabfit}.

Another investigation performed in this work was the analysis of how the nuclear matter bulk parameter, namely, incompressibility, effective mass, symmetry energy and its slope, affect the crustal properties of the neutron star. We verify that the symmetry energies is the quantity that produces the higher variation in $M_{\rm crust}$, $R_{\rm crust}$, and $\Delta I/I$, according to the results presented in Figs.~\ref{mcrust-corr}, \ref{rcrust-corr}, and~\ref{deltai-corr}. Furthermore, we were able to construct a particular parametrization in which the $\Delta I/I \geqslant 7\%$ constraint is satisfied for neutron stars masses of $M=1.4M_\odot$. This particular value of $M$ was used in Ref.~\cite{pavlov} to properly fit data from the softer component of the Vela pulsar X-ray spectrum. The bulk parameters found for this purpose are $\rho_0=0.15$~fm$^{-3}$, $B_0=-16.0$~MeV, $m^*=0.575$, $K_0=260$~MeV, $J=35$~MeV, and $L_0=70$~MeV for the model with $\alpha_1=\alpha'_1=\alpha_2=\alpha'_2=0$. The mass-radius diagram for this specific RMF parametrization, displayed in Fig.~\ref{mr-new}, shows that it is compatible with data from pulsars PSR J1614-2230, $M=1.97\pm 0.04M_{\odot}$~\cite{Demorest}, PSR J0348+0432, $M=2.01\pm 0.04M_{\odot}$~\cite{Antoniadis}, and MSP J0740+6620, $M=2.14{{+0.20}\atop{-0.18}}M_{\odot}$~\cite{Cromartie}. We also verify agreement with data from the NICER mission, namely, $M=1.44^{+0.15}_{-0.14}M_{\odot}$ with $R=13.02^{+1.24}_{-1.06}$~km~\cite{Col}, $M=1.34^{+0.15}_{-0.16}M_{\odot}$ with $R=12.71^{+1.14}_{-1.19}$~km~\cite{l21}, and $R_{1.44}>10.7$~km~\cite{l25}. The radius for the Vela pulsar with entrainment effects included was estimated to satisfy $R \geqslant 3.24293 + 4.43579(M/M_\odot) - 0.39817(M/M_\odot)^2$.

\section*{Appendix}
\label{app}

Here we derive the mainly equations of the approach used for the crustal star properties, based on the core equation of state and macroscopic properties of the core-crust interface~\cite{AA119-599}. This formalism is based on the assumption that the mass of the crust, $M_{\rm crust}$, is much smaller than the total mass of the star, $M$. We start by constructing the Tolman–Oppenheimer–Volkoff equations.

In order to describe a spherically symmetric object, we use the metric given by
\begin{equation}
ds^2 = e^{\varphi(r)} \, c^2 \, dt^2 -  e^{\xi(r)} dr^2 - r^2 \left(d\theta^2 + \sin^2\theta d\phi^2\right).
\end{equation}
We consider the stress-energy tensor for a perfect fluid as $T_{\mu\nu} = \left(P/c^2+\rho \right) u_{\mu}\, u_{\nu} - g_{\mu\nu} \, P/c^2 $, where $P$ is the pressure and $\mathcal{E}$ is energy density. The quadrivelocity $u_{\mu}$ fulfills $u_{\mu} \, u^{\mu}=1$ and $\, u^{\mu} \nabla_{\nu }u_{\mu} =0$. By defining 
\begin{eqnarray}
e^{-\xi} = \left[1 - \frac{2 G m(r)}{r c^2}\right], 
\end{eqnarray}
where $m=m(r)$ represents the mass inside a sphere of radius $r$, we obtain the system of equations that governs the hydrostatic equilibrium of the compact star, namely,
\begin{align}
m'(r) &= 4 \pi  \mathcal{E}(r) \, r^2,  \\
P'(r) &= 
\nonumber\\
&- \frac{\left[P(r)/c^2+ \mathcal{E}(r) \right] \left[ \frac{4 \pi G}{c^2}\, r \, P(r) + \frac{G \, m(r)}{r^2}\right]} {\left[1 - \frac{2 \, G\,  m(r)}{r c^2}\right]} \, . 
\label{pla} 
\end{align}
In the limit $M_{\rm crust} \ll M$, we have $4\pi r^3 P(r)/[m(r)c^2] \ll 1$. Therefore, for the crust we can write Eq.~(\ref{pla}) as
\begin{eqnarray}
\frac{P'(r)}{P(r) + \mathcal{E}(r)\, c^2} = - \frac{G \, m(r)}{c^2 \, r^2} \left[1 - \frac{2 \, G\,  m(r)}{r c^2}\right]^{-1} \, . 
\label{eq22}
\end{eqnarray}

We can estimate the contribution of the crust mass by considering an approximation where $P(r)/c^2$ is much smaller than $\mathcal{E}(r)$. In the crust region, for $M_{\rm crust} \ll M$ and using $dm = \mathcal{E}(r) dv = \mathcal{E}(r) 4\pi r^2 dr$, we obtain

\begin{eqnarray}
\frac{dP(r)}{dm} = - \frac{G \, M}{4\pi \, r^4} \left[1 - \frac{2 \, G\,  M}{r c^2}\right]^{-1} \, .
\label{eq23}
\end{eqnarray}

Integrating Eq. (\ref{eq23}) from the outer crust up to the core-crust interface, we have

\begin{eqnarray}
M_{\rm crust} = \frac{4\pi P_t R^4_{\rm core}}{GM_{\rm core}}\left(1-\frac{2GM_{\rm core}}{R_{\rm core}c^2}\right)\, .
\label{mcrusta}
\end{eqnarray}

On the other hand, one can use the thermodynamic relation $\mu=d\mathcal{E}/d\rho$ and
$\mu = \left(P(r) + \mathcal{E}(r)\, c^2\right)/\rho $
to write
\begin{eqnarray}
\frac{d\mu}{\mu} = \frac{dP}{P(r) + \mathcal{E}(r)\, c^2} \, ,
\end{eqnarray}
which is precisely the left hand side of Eq. (\ref{eq22}). Combining those expressions and using $M_{\rm core} \ll M$, one obtains, in the crust region,  the following:
\begin{eqnarray}
\frac{d\mu}{\mu} = - dr \, \frac{G \, M}{c^2 \, r^2} \left[1 - \frac{2 \, G\,  M}{r c^2}\right]^{-1} \, . 
\label{eq26}
\end{eqnarray}
Integrating from the outer crust up to the core-crust transition, we have
\begin{eqnarray}
\int_{\mu_0}^{\mu_t} \frac{d\mu}{\mu} &=& 
- \int_{R_{\rm core}}^{R}  \frac{G \, M}{c^2 \, r^2} \left[1 - \frac{2 \, G\,  M}{r c^2}\right]^{-1}  \, dr \, , \\
\left(\frac{\mu_t}{\mu_0}\right)^2 &=& \dfrac{1-\dfrac{2 \, G\, M}{R\, c^2}}{1- \dfrac{2 \, G\, M}{R_{\rm core}\, c^2}} \, ,
\label{eq28}
\end{eqnarray}
where $\mu_t$ is the baryon chemical potential at the core-crust transition and $\mu_0$ is the chemical potential at the surface of the neutron star.

Using Eq.({\ref{eq28}}), one can obtain 
\begin{eqnarray}
 R = \frac{2 \, G \, M \, R_{\rm core}}{c^2 \, R_{\rm core} + 2 \, G \, M \, \left(\dfrac{\mu_t}{\mu_0}\right)^2 - c^2 \, R_{\rm core} \left(\dfrac{\mu_t}{\mu_0}\right)^2}\, .
\end{eqnarray}
Finally, the crustal radius is~\cite{AA119-599}
\begin{align}
&R_{\rm crust} = R - R_{\rm core}
\nonumber\\
&= \frac{2 \, G \, M \, R_{\rm core}}{c^2 \, R_{\rm core} + 2 \, G \, M \, \left(\dfrac{\mu_t}{\mu_0}\right)^2 - c^2 \, R_{core} \left(\dfrac{\mu_t}{\mu_0}\right)^2} 
- R_{core}
\nonumber\\
&= \phi R_{\rm core} \left[\frac{1-R_{\rm s}/R_{\rm core}}{1-\phi\left(1-R_{\rm s}/R_{\rm core}\right)}\right],
\label{rcrusta}
\end{align}
with $R_{\rm s} = 2GM/c^2$ and $\phi = [(\mu_t/\mu_0)^2 - 1]R_{\rm core}/R_{\rm s}$.

\section*{ACKNOWLEDGMENTS}
This work is a part of the project INCT-FNA proc. No. 464898/2014-5. It is also supported by Conselho Nacional de Desenvolvimento Científico e Tecnológico (CNPq) under Grants No. 310242/2017-7, 312410/2020-4, 406958/2018-1 (O.L.), No. 433369/2018-3 (M.D.), and 438562/2018-6, 313236/2018-6 (W.P.). We acknowledge the partial support of CAPES under Grant No. 88881.309870/2018-01 (W.P.). We also acknowledge Funda\c{c}\~ao de Amparo \`a Pesquisa do Estado de S\~ao Paulo (FAPESP) under Thematic Project 2017/05660-0 and Grant No. 2020/05238-9 (O.L., M.D., C.H.L).

\section*{Data Availability Statement}

This manuscript has no associated data
or the data will not be deposited. [Authors’ comment: All data generated
during this study are contained in this published article.]
\\

\end{document}